\documentclass[preprint,preprintnumbers,amsmath,amssymb, nofootinbib]{revtex4}
\pdfoutput=1 

\usepackage{graphicx}
\usepackage{dcolumn}
\usepackage{bm}
\usepackage{epsfig}
\usepackage{color}
\usepackage{multirow}

\usepackage{amsmath}
\usepackage{amssymb}

\def\fun#1#2{\lower3.6pt\vbox{\baselineskip0pt\lineskip.9pt
  \ialign{$\mathsurround=0pt#1\hfil##\hfil$\crcr#2\crcr\sim\crcr}}}

\input epsf

\newcommand{\be}{\begin{equation}}
\newcommand{\ee}{\end{equation}}
\newcommand{\bea}{\begin{eqnarray}}
\newcommand{\eea}{\end{eqnarray}}

\begin{document}

\vspace*{2cm}

\title{Improving the sensitivity of Higgs boson searches \\ 
in the golden channel}


\author{
\vspace{0.5cm} 
James S. Gainer$\,^{a,b}$, Kunal Kumar$\, ^{b}$, Ian Low$\, ^{a,b}$, and  
Roberto Vega-Morales$\, ^{b}$ }

\affiliation{
\vspace*{.2cm}
$^a$ \mbox{High Energy Physics Division, Argonne National Laboratory, 
Argonne, IL 60439}\\
$^b$ \mbox{Department of Physics and Astronomy, Northwestern University, 
Evanston, IL 60208} \\
\vspace*{0.8cm}}

\begin{abstract}
\vspace*{0.5cm}
Leptonic decays of the Higgs boson in the $ZZ^{(*)}$ channel yield what is
known as the golden channel due to its clean signature and good total 
invariant mass resolution. 
In addition, the full kinematic distribution of the decay products 
can be reconstructed, which, nonetheless, is not taken into account in 
traditional search strategy relying only on measurements of the total 
invariant mass. In this work we implement a type of multivariate analysis
 known as the matrix element method, which exploits differences in the full 
production and decay matrix elements between the Higgs boson and the dominant 
irreducible background from $q\bar{q}\to ZZ^{(*)}$. Analytic expressions of 
the differential distributions for both the signal and the background are 
also presented. 
We perform a study for the Large Hadron Collider  at $\sqrt{s}=7$ 
TeV for Higgs masses between $175$ and $350$ GeV. 
We find that, with an integrated luminosity of 2.5 fb$^{-1}$ or higher, 
improvements in the order of $10 - 20 \%$ could be obtained 
for both discovery significance and exclusion 
limits in the high mass region, where 
the differences in the angular correlations 
between signal and background are most pronounced.
\end{abstract}


\maketitle

\section{Introduction}\label{introduction}

The discovery of the Higgs boson~\cite{Djouadi:2005gi} would be the 
triumphant culmination of the experimental quest to discover the particles 
of the Standard Model.
The Tevatron has already set interesting limits on the Standard Model (SM) 
Higgs boson in the intermediate mass range~\cite{CDF:2011cb}. 
At the Large Hadron Collider (LHC), with data corresponding to 
roughly $1$ fb$^{-1}$ of integrated luminosity,  
the ATLAS collaboration has announced exclusions at $95\%$ confidence level 
of  Higgs masses in the ranges $155 - 190$ GeV and 
$295 - 450$ GeV~\cite{ATLAS-CONF-2011-112}, while the 
CMS collaboration's limits are in the ranges $149 - 206$ GeV 
and  $300 - 440$ GeV~\cite{CMS-PAS-HIG-11-011}. 
High mass limits from both collaborations are driven by measurements 
in the $ZZ^{(\ast)}$ channel, which is considered the main discovery channel 
of the Higgs boson for masses above $200$ GeV.

Among the different decay products of the Higgs into $ZZ^{(\ast)}$ bosons, 
the one with both $Z$ bosons decaying into $e^+e^-$ or $\mu^+\mu^-$ is 
often referred to as ``the golden channel'' because of the good invariant mass 
resolution and well-controlled background. The traditional search strategy 
using the golden channel thus focuses on measuring the invariant mass spectrum 
of the four leptons. However, given that four-momenta of all decay products 
can be reconstructed with sufficient resolution, 
it is possible to measure more than just the total invariant mass of the four 
leptons. In fact, there are a total of five angles (and two 
additional invariant masses, those of the off-shell $Z$ bosons) 
that can be measured. Obviously it would be 
advantageous to incorporate all available kinematic information when 
searching for the Higgs boson.

Additional kinematic variables can be included in an experimental measurement 
by multivariate analyses~\cite{Bhat:2010zz}, 
which already have a wide range of applications in many 
measurements done at the Tevatron and the $B$ factories. 
Several multivariate methods have been employed, 
such as neural nets;  boosted decision trees;  
and the Matrix Element Method (MEM)~\cite{Kondo:1988yd}, 
the best-known use of which has been in studying the top quark at the 
Tevatron~\cite{Abbott:1998dn}.
Even in some of the LHC Higgs analyses, 
most notably leptonic decays of Higgs to $W^+W^-$ where there are two missing 
neutrinos, multivariate analyses, in particular boosted decision trees, 
have been used to incorporate additional kinematic observables 
such as the opening angle between the two charged 
leptons in the final states~\cite{CMS-PAS-HIG-11-003}.  
On the other hand, it is somewhat surprising that in the golden channel, 
where there is no missing 
particle in the final state and all angles can be reconstructed, no 
experimental analysis that we are aware of has considered supplementing 
total invariant mass with angular correlations to search for the Higgs boson.
(For recent analyses, see~\cite{ATLAS-CONF-2011-112,CMS-PAS-HIG-11-004}.)

Angular correlations of Higgs decays in the golden channel have been studied 
previously, to determine the spin and CP properties of the putative Higgs 
resonance~\cite{Matsuura:1991pj,Buszello:2002uu}. 
A particularly useful observable, the azimuthal angle between the 
decay planes of the two $Z$ bosons, was pointed out recently in 
Refs.~\cite{Keung:2008ve,Cao:2009ah}. 
Subsequently, two comprehensive studies appeared in 
Refs.~\cite{Gao:2010qx,DeRujula:2010ys}.  These works 
included the computation of the angular correlations of the 
final state leptons
resulting from the production of a resonance (with arbitrary spin
less than or equal to two) which in turn decays, via general couplings, 
to a pair of $Z$ bosons, which subsequently decay leptonically.
Both analyses also implemented the MEM in this $ZZ^{(\ast)}\to 4\ell$ 
channel, 
to distinguish between various hypotheses for the spin and CP properties 
of the putative Higgs signal at the LHC with $\sqrt{s}= 14$ and $10$ TeV, 
respectively. 
Moreover, Ref.~\cite{DeRujula:2010ys} briefly discussed using the 
MEM to enhance the Higgs discovery reach in the golden channel 
at $10$ TeV for two specific values of the Higgs mass ($200$ and $350$ GeV).

In the present work, instead of comparing angular correlations for different 
spin and CP assumptions for a singly produced resonance, we aim at 
distinguishing the SM Higgs boson signal from the dominant irreducible 
background $q\bar{q}\to ZZ^{(\ast)} \to 4\ell$ using the MEM at the LHC with 
$\sqrt{s}=7$ TeV. 
The scattering amplitude for $q\bar{q}\to ZZ^{(\ast)} \to 4\ell$, 
when both $Z$ bosons are on-shell, has been computed long ago in 
Refs.~\cite{Gunion:1985mc, Hagiwara:1986vm}. 
We extend the calculation of Ref.~\cite{Hagiwara:1986vm} 
to off-shell $Z$ bosons and present analytic 
expressions for the fully differential cross section. Then we perform a 
Monte Carlo study, implementing the MEM to determine the improvement
in sensitivity for Higgs boson searches 
in the golden channel over a significant range of 
Higgs masses (between $175$ and $350$ GeV).
We perform these analyses for several integrated luminosities, 
between $1$ and $7.5$ fb$^{-1}$. For simplicity we consider only the 
0-jet bin and assume that events have no intrinsic
$p_T$, though we expect the qualitative features of our results to be
more generally applicable.

This work is organized as follows: in Sect.~\ref{kinematics} we introduce 
and define kinematic variables to be used in the fully differential 
cross sections. In particular, Lorentz-invariant expressions for all 
production and decay angles are presented, so that the kinematic distributions 
can be reconstructed using measurements done in the laboratory frame. 
In Sect.~\ref{diffx} we compute the amplitude and cross section for 
$q\bar{q}\to ZZ^{(\ast)}\to 4\ell$ using the technique of helicity amplitudes 
introduced in Ref.~\cite{Hagiwara:1985yu}, allowing both $Z$ bosons to be 
off-shell. In Sect.~\ref{statistics} the MEM is briefly reviewed, as well 
as the relevant statistical procedures we employed. The Monte Carlo study, 
including event generation, detector smearing effects, and construction of 
pseudo-experiments is discussed in Sect.~\ref{angular2}. Then we present 
our results for both expected significance and exclusion limits in 
Sect.~\ref{results}. Finally, we close with conclusions in 
Sect.~\ref{conclusions}.

\section{Kinematics}\label{kinematics}

As noted above, in this study 
we consider events in which two $Z$ bosons are produced, 
either from the decay of a SM Higgs Boson produced in the 
gluon fusion channel or from $t$($u$)-channel $q\bar{q}$ production.  
Each $Z$ boson, which could be either on or off the mass shell,
decays to a lepton ($\ell$) and an anti-lepton (${\bar{\ell}}$).  
We do not consider events with additional particles in the final state; 
thus the transverse momentum of the $4\ell$ system is assumed to be negligible.
In other words, we only consider exclusive $ZZ^{(\ast)}\to 4\ell$ processes.

In these events, the final state can be completely reconstructed. 
In general the kinematics can be specified in terms of two production 
angles of the $ZZ^{(\ast)}$ system, one of which is irrelevant; 
four decay angles describing $ZZ^{(\ast)}\to 4 \ell$; 
and the invariant masses of the $Z$'s.  
In hadron colliders it is also necessary to know the momentum fractions 
of the initial massless partons, $x_1$, and $x_2$, 
in order to  compute the differential cross sections.
We now describe our convention for the angles which specify the event 
and how to obtain them for a particular event. 
In particular, we provide Lorentz-invariant definitions of all angles, 
allowing for their determination from four-momenta reconstructed in 
the laboratory (Lab) frame.

\subsection{Definition of angles}

Let $p_1$ and $p_2$ be the momenta of the lepton pair coming from $Z_1$, 
and $p_3$ and $p_4$ be the momenta of the lepton pair from $Z_2$, 
while $k_{1,2}$ are the momenta of $Z_{1,2}$. Our notation is such that 
$p_1=\ell_1$, $p_2=\bar\ell_1$, $p_3=\ell_2$, $p_4=\bar\ell_2$, 
i.e. $p_1$ is the momentum of the lepton from $Z_1$ decay, 
$p_2$ the momentum of the antilepton from $Z_1$ decay, etc. 
We denote the momenta of the incoming partons by  $k_q$
 and $k_{\bar{q}}$.
The total momentum of the $ZZ^{(\ast)}$ system is 
$P=k_q+k_{\bar{q}}=k_1+k_2=p_1+p_2+p_3+p_4$, which satisfies 
$P^2=\hat{s}\equiv M^2$. For Higgs production in the 
gluon fusion channel, the incoming partons are self-conjugate, 
$k_q=k_{\rm{gluon},1}$, $k_{\bar{q}}=k_{\rm{gluon},2}$, 
and the total momentum $P$ is the Higgs momentum.  

\begin{figure}
\includegraphics[scale=0.85, angle=0]{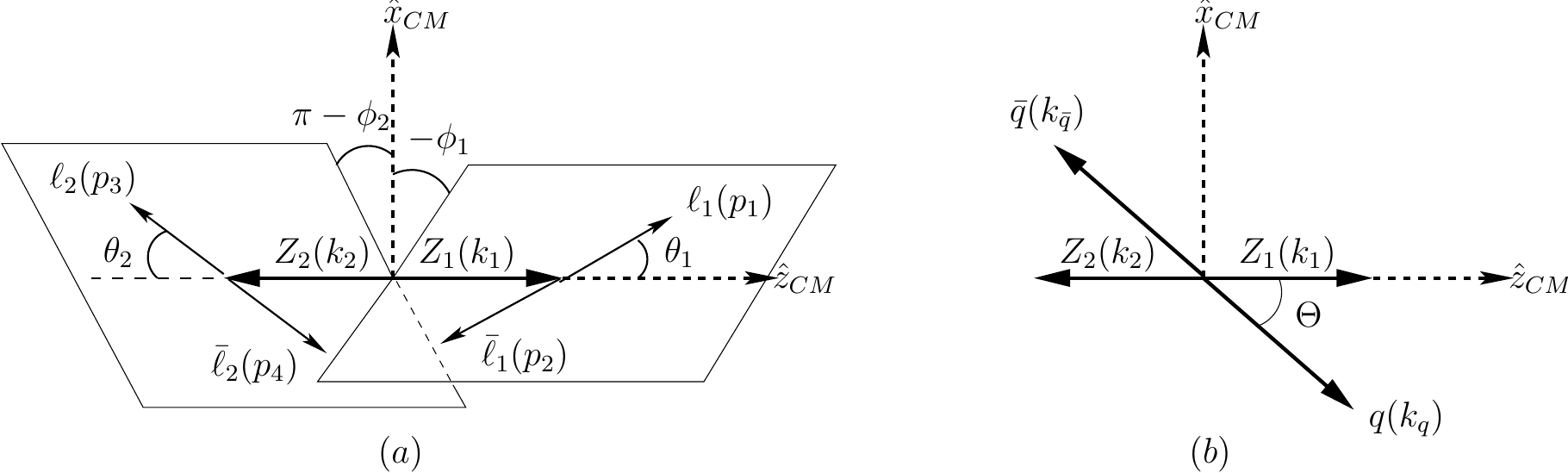}  
\caption{\label{fig1}{\em (a) Two decay planes of 
$Z_i\to  \ell_i\bar\ell_i$, $i=1,2$. The polar angles 
$\theta_{i}$ shown are defined in the rest frames of 
$Z_{i}$ with respect to $\hat{k}_{i}$, while the azimuthal angles shown 
are in fact $2\pi-\phi_{1} = -\phi_{1}$ and $\pi-\phi_2$.  
(b) The coordinate system in the CM frame and the definition of the 
production angle $\Theta$.
}}
\end{figure}

As indicated in Fig.~\ref{fig1}, we choose the coordinate system in the 
center-of-mass (CM) frame of the two $Z$'s system  as:
\be
\hat{z}_{CM} = \hat{k}_1 \ , \qquad \hat{y}_{CM} = 
\frac{\hat{k}_q \times \hat{k}_1}{|\hat{k}_q\times \hat{k}_1|} \ , 
\qquad \hat{x}_{CM}=\hat{y}_{CM}\times\hat{z}_{CM}
 =\frac{-\hat{k}_q +
\hat{k}_1 (\hat{k}_q\cdot\hat{k}_1)}{|\hat{k}_q\times \hat{k}_1|}\ .
\ee 
Furthermore, we define ${\cal Z}_1$ as the rest frame of the $Z_1$ boson by 
boosting the CM frame along $\hat{k}_1$, while ${\cal Z}_2$ is obtained by
 first rotating CM frame with respect to $\hat{y}_{CM}$ by $\pi$ and then 
boosting along $\hat{k}_2$. The production angle $\Theta$ and decay angles 
$\{\theta_1,\theta_2,\phi_1,\phi_2\}$ are defined as follows:
\begin{itemize}

\item $\Theta$: polar angle of the momentum of the incoming quark in the 
CM frame.

\item $\theta_{1,2}$: polar angle of the momentum of $\ell_{1,2}$ in the 
${\cal Z}_{1,2}$ frame.

\item $\phi_{1,2}$: azimuthal angle of $\ell_{1,2}$ in the ${\cal Z}_{1,2}$ 
frame.

\end{itemize}
The azimuthal production angle  is irrelevant and chosen to be zero. 
In these definitions, three-momenta of $\ell_{1,2}$ in the  
${\cal Z}_{1,2}$ frame can be written as
\be
\vec{p}_{\ell_i}\  \  \mbox{in the ${\cal Z}_i$ frame} 
=|\vec{p}_{\ell_i} |\ (\sin{\theta_i}\cos{\phi_i},  
\sin{\theta_i} \sin{\phi_i}, \cos{\theta_i}) \ ,\  i = 1, 2 \ , 
\ee
while the three-momentum of the incoming parton in the CM frame is
\be
\vec{k}_q\  \ \mbox{in the CM frame} = |\vec{k}_q| \ 
(-\sin \Theta, 0, \cos\Theta) \ .
\ee

In hadron colliders the CM frame of the two $Z$'s system is different from 
the Lab frame and the event as a whole will be boosted along the beam axis 
with respect to the Lab frame, $P=(P^0, 0,0, P^z)$.  
Also, we have chosen to define the coordinate system in the CM frame 
such that the $\hat{z}$ axis is defined by the $Z_1$ three-momentum,
rather than by the three-momentum of the incident partons, as is
natural in the Lab frame.

The total energy and 
momentum of the event $P$ in the Lab frame can be used to determine the 
momentum fractions of the  incident partons.   Following 
Ref.~\cite{Peskin:1995ev}, we write $k_q=x_1(E_{cm}, 0,0, E_{cm})$ and 
$k_{\bar{q}}=x_2(E_{cm}, 0,0, -E_{cm})$, where $E_{cm}=\sqrt{s}/2$ is 
the CM energy of the colliding protons. From $P=k_q+k_{\bar{q}}$ we see 
that $\hat{s}=x_1 x_2 s$ and
\bea
k_q&=&\frac12( P^0+P^z,0,0,P^0+P^z) \ , \\
 k_{\bar{q}}&=&\frac12 ( P^0-P^z,0,0,P^z-P^0)\ ,
\eea
which are valid in the Lab frame.

\subsection{Lorentz-invariant construction of angles}

In the CM frame, $\vec{k}_1$ and $\vec{k}_2$ are back to back and of equal 
magnitude, as are $\vec{k}_q$ and $\vec{k}_{\bar{q}}$. Using $P=M(1,0,0,0)$ 
we can work out the energy and three-momentum of the incoming partons,
\be
E_q=E_{\bar{q}}=|\vec{k}_q|=|\vec{k}_{\bar{q}}|=\frac{\sqrt{\hat{s}}}2 \ , 
\ee
as well as that of the two $Z$'s,
\be
\label{eq:Eiki}
E_i=\frac{P\cdot k_i}{M} \ , \quad |\vec{k}_i|=
\sqrt{\left(\frac{P\cdot k_1}{M}\right)^2-m_{12}^2}  \equiv \lambda_Z \ , 
\quad  i=1,2 \ ,
\ee 
where we define $m_{ij}^2=(p_i+p_j)^2=2p_i \cdot p_j$. 
Alternatively, $\lambda_Z=\sqrt{\left(P\cdot k_2/M\right)^2-m_{34}^2}$.

Since $\cos\Theta=\hat{k}_q\cdot \hat{k}_1$, by computing $k_q\cdot k_1$ 
it is simple to derive
\be
\label{eq:cosTh}
\cos \Theta = \frac{-k_q\cdot k_1+E_q E_1}{|\vec{k}_q||\vec{k}_1|} 
=\frac{(k_{\bar{q}}-k_q)\cdot k_1}{M \ \lambda_Z} \ .
\ee
By definition $\cos\Theta$ changes sign under 
$\hat{k}_q\leftrightarrow \hat{k}_{\bar{q}}$, 
which is manifest in Eq.~(\ref{eq:cosTh}). 
Thus when the direction of the incoming quark cannot 
be distinguished from the anti-quark, as is the case for hadron colliders, 
or when the incoming partons are self-conjugate as in the 
Higgs production channel we consider, one can only determine 
$\cos\Theta$ up to a minus sign. 
Because $\Theta$ is only defined between 0 and $\pi$, 
it is not necessary to compute $\sin\Theta$.

Next we consider $\theta_i$, which was worked out in the 
appendix of Ref.~\cite{Cao:2009ah} . Since $k_1=p_1+p_2$, 
in the CM frame  we can write $k_1=(E_1+E_2,0,0,|\vec{p}_1+\vec{p}_2|)$
and solve for the boost that takes $k_1$ from the CM frame to the
 rest frame of $Z_1$ where it is $(m_{12},0,0,0)$:
\be
\left( \begin{array}{c}
                m_{12} \\
                0 
                \end{array} \right) =
     \left(\begin{array}{cc}
                \gamma & -\gamma\beta \\
                -\gamma\beta & \gamma 
                \end{array}\right)  
  \left( \begin{array}{c}
                E_1+E_2 \\
                |\vec{p}_1+\vec{p}_2| 
                \end{array} \right)        \ ,
\ee        
from which we get
\be
\beta = \frac{  |\vec{p}_1+\vec{p}_2| }{E_1+E_2} \ , 
\qquad \gamma = \frac{E_1+E_2}{m_{12} }\ .
\ee   
The inverse boost would then take $p_1=(m_{12}/2)(1,\sin\theta_1 \cos\phi_1, 
\sin\theta_1 \sin\phi_1,\cos\theta_1)$ in the $Z_1$ rest frame to 
$p_1=(E_1,\vec{p}_1)$ in the CM frame, implying the following relation:
\be
E_1 = \gamma \frac{m_{12}}2 (1+\beta \cos\theta_1) \ ,
\ee 
from which we obtain 
\be
\cos\theta_1 = \frac{E_1-E_2}{ |\vec{p}_1+\vec{p}_2|} =  
\frac{E_1-E_2}{ |\vec{k}_1|}\ .
\ee
Using Eq.~(\ref{eq:Eiki}) we arrive at 
\be
\cos\theta_1= \frac1{M \lambda_Z} P\cdot(p_1-p_2)\ .
\ee
For $\cos\theta_2$, simply replace $p_1$ and $p_2$ by $p_3$ and $p_4$, 
respectively, and we obtain
\be
\cos\theta_2= \frac1{M \lambda_Z} P\cdot(p_3-p_4) \ .
\ee

To compute $\phi_{i}$, we first construct the unit normal vectors of 
the two decay planes,
\be
\hat{N}_1 = \frac{\vec{p}_1\times \vec{p}_2}{|{\vec{p}_1\times \vec{p}_2}|} \ 
, \qquad
\hat{N}_2 = \frac{\vec{p}_3\times \vec{p}_4}{|{\vec{p}_3\times \vec{p}_4}|} \ ,
\ee
so that  
\bea
&& \hat{N}_1\cdot \hat{x}_{CM}  = \sin \phi_1\ , \qquad  \, 
\hat{N}_1\cdot \hat{y}_{CM}=-\cos\phi_1 \ , \\
&& \hat{N}_2\cdot \hat{x}_{CM}  = -\sin \phi_2\ , \qquad  \, 
\hat{N}_2\cdot \hat{y}_{CM}=-\cos\phi_2 \ . 
\eea 
The numerator of the normal vector can be written as
\be
(\vec{p}_i\times \vec{p}_j)^a = \frac1{M}\epsilon^{\mu\nu a\rho} 
p_{i\mu}p_{j\nu}P_\rho\equiv \frac1{M}\epsilon^{p_i p_j a P} \ ,
\ee
where $\epsilon_{0123}=-\epsilon^{0123}=1$. On the other hand, 
the Lorentz-invariant form of the numerator can be obtained using the relations
\be 
|\vec{p}_i|=\frac1M \, p_i\cdot P \ ,\quad 
\cos{\bar\theta}_{ij}=1-{m_{ij}^2\over  2|\vec{p}_i| |\vec{p}_j|}  \ ,
\ee
where $\bar{\theta}_{ij}$ is the opening angle between $\vec{p}_i$ and 
$\vec{p}_j$ in the CM frame, so that in the end we have
\be
|\vec{p}_i\times \vec{p}_j|=|\vec{p}_i| |\vec{p}_j| \sin\bar{\theta}_{ij}=
m_{ij} \left(\frac{p_i\cdot P}{M} \frac{p_j\cdot P}{M}-
\frac{m_{ij}^2}4\right)^{\frac12} \equiv \kappa_{ij} \ .
\ee 
One then calculates
\bea
\sin\phi_1 &=& -\frac{\vec{p}_1\times \vec{p}_2}{|{\vec{p}_1\times 
\vec{p}_2}|} \cdot \frac{\hat{k}_q}{\sin\Theta}
 = \frac{2}{M^2 \kappa_{12}  \sin\Theta}  \epsilon^{p_1p_2k_qP} \ , \\
 \cos \phi_1 &=& - \frac{\vec{p}_1\times \vec{p}_2}{|{\vec{p}_1\times 
\vec{p}_2}|} \cdot  \frac{\hat{k}_q\times 
\hat{k}_1}{|{\hat{k}_q\times \hat{k}_1}|} =
 -\frac{2}{M^3\kappa_{12}  \lambda_Z \sin\Theta} 
  \left|
\begin{array}{ccc}
p_1\cdot k_q & \ p_1\cdot k_1 \ & p_1\cdot P\\
p_2\cdot k_q & \ p_2\cdot k_1 \ & p_2\cdot P\\
P\cdot k_q  & \ P\cdot k_1   \ &   M^2 \end{array}\right| \ ,
\eea 
and similarly
\bea
\sin\phi_2 &=& -\frac{2}{M^2\kappa_{34} \sin\Theta}  
\epsilon^{p_3p_4k_qP} \ , \\
 \cos \phi_2 &=&  -\frac{2}{M^3\kappa_{34}  \lambda_Z \sin\Theta} 
  \left|
\begin{array}{ccc}
p_3\cdot k_q & \ p_3\cdot k_1 \ & p_3\cdot P\\
p_4\cdot k_q & \ p_4\cdot k_1 \ & p_4\cdot P\\
P  \cdot k_q & \ P\cdot k_1   \ &   M^2 \end{array}\right| \ .
\eea 
It is also worth noting that when $\hat{k}_q\to -\hat{k}_q$, 
$\phi_i \to \pi+\phi_i$. So in hadron colliders or gluon fusion 
production we cannot distinguish 
between an event described by angles 
($\Theta,\theta_1,\theta_2,\phi_1,\phi_2$) and an event described
by angles ($\pi - \Theta, \theta_1, \theta_2, \phi_1 + \pi, \phi_2 + \pi$).

\section{Differential Cross Sections}\label{diffx}

Angular distributions in $ZZ^{(\ast)} \to 4\ell$ provide a wealth of 
information on the production mechanism of the two $Z$ 
bosons~\cite{Buszello:2002uu,Keung:2008ve, Cao:2009ah, 
Gao:2010qx, DeRujula:2010ys, Kniehl:1990mq}. 
Similar angular correlations in the vector boson fusion channel of Higgs 
production have also been discussed in 
Ref.~\cite{Hankele:2006ma}.
As noted above, 
in this work we focus on the search of the Higgs boson in the golden channel, 
$h\to ZZ^{(\ast)} \to 4\ell$, and study the possibility of differentiating 
the Higgs signal from the dominant irreducible background 
$q\bar{q}\to ZZ^{(\ast)}\to 4\ell$ using spin correlations.  
In particular, we will compute the amplitudes in a helicity basis 
following Ref.~\cite{Hagiwara:1986vm}. 

We will present the expressions for the fully differential cross section 
for both the signal and the background. Results for the Higgs production 
and decay have appeared in many previous works (see, for example, 
Refs.~\cite{Buszello:2002uu,Cao:2009ah,Gao:2010qx,DeRujula:2010ys,Low:2010jp})
and are not new. They are given here for completeness. 
Earlier works on $q\bar{q}\to ZZ^{(\ast)} \to 4\ell$ 
include~\cite{Gunion:1985mc, Hagiwara:1986vm}.

\begin{figure}
\includegraphics[scale=0.85, angle=0]{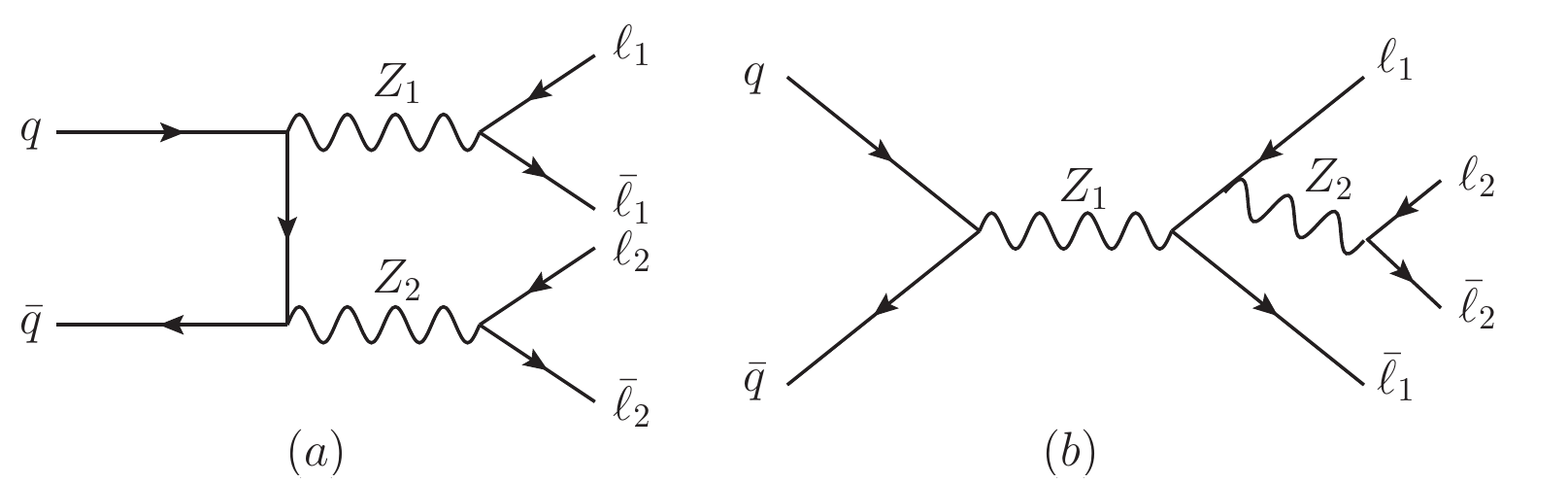}  
\caption{\label{fig2}{\em Feynman diagrams contributing to 
$q\bar{q}\to 4\ell$. We only consider (a), since final states 
from (b) have a total invariant mass at the $Z$ mass.
}}
\end{figure}

Feynman diagrams contributing to $q\bar{q}\to 4\ell$ are 
shown in Fig.~\ref{fig2}.
We consider only  the diagram in (a) and the 
corresponding $u$-channel diagram, as we are interested 
in final states with a total invariant mass much larger than the $Z$ mass.
We will compute the amplitude of the diagram in Fig.~\ref{fig2} (a) and 
its $u$-channel partner in a helicity basis, following 
Ref.~\cite{Hagiwara:1986vm}. The amplitude for the process under consideration 
factorizes into one production and two decay amplitudes:
\bea
\label{proc:qqzz}
q(k_q, \sigma) + \bar{q}(k_{\bar{q}},\bar{\sigma})& \longrightarrow & 
Z_1 (k_1,\lambda_1) + Z_2 (k_2,\lambda_2)~, \\
\label{proc:z1ll}
 Z_1(k_1,\lambda_1) &\longrightarrow & \ell_1(p_1,\sigma_1) + 
\bar{\ell}_1(p_2,\sigma_2)~,  \\
 Z_2(k_2,\lambda_2) &\longrightarrow & \ell_2(p_3,\sigma_3) + 
\bar{\ell}_2(p_4,\sigma_4)~,  
 \eea
where the momentum and helicity of each particle are indicated. 
A similar factorization obtains in the case of the 
$gg \to h \to ZZ^{(\ast)} \to 4\ell$ signal.
As the production and decay amplitudes factorize, 
we will consider them separately.  


Before going into details of the computation, we state here the conventions we choose for various kinematic vectors and fermion spinors. We  use the explicit
expressions for the four-momenta and polarization vectors of the 
$Z$ bosons in the CM frame:
\bea\label{definitions}
k_1 &=& m_1 \gamma_1 (1,0,0,\beta_1) \ , \\
 k_2 &=& m_2 \gamma_2 (1,0,0,-\beta_2)   \ , \\
 \epsilon_1^{(\pm)}&=& \frac1{\sqrt{2}}(0,\mp 1,i,0)\ , 
\quad \epsilon_1^{(0)}=\gamma_1(\beta_1,0,0,1) \ , \\
  \epsilon_2^{(\pm)}&=& \frac1{\sqrt{2}}(0,\pm 1, i,0)\ , 
\quad \epsilon_2^{(0)}=\gamma_2(\beta_2,0,0,-1) \ ,
 \eea
 where $k_i^2=m_i^2, i=1,2$ is the invariant mass of $Z_i$, which could 
be off the mass shell, and the boost factors are
 \be
 \gamma_1 =\frac1{\sqrt{1-\beta_1^2}}= \frac{\sqrt{\hat{s}}}{2m_1}(1+x)  \ , 
\quad \gamma_2=\frac1{\sqrt{1-\beta_2^2}} = \frac{\sqrt{\hat{s}}}{2m_1}  
(1-x) \ , \quad x = \frac{m_1^2-m_2^2}{\hat{s}} \ .  
 \ee
We will also use the explicit forms of $u(p,\lambda)$ and 
$v(p,\lambda)$ spinors:
\bea\label{spinor definitions}
u_R = \begin{pmatrix} 0 \\ 0 \\ \sqrt{2E} \\ 0 \end{pmatrix},~
u_L = \begin{pmatrix} 0 \\ \sqrt{2E} \\ 0 \\ 0 \end{pmatrix},~
v_R= \begin{pmatrix} \sqrt{2E} \\ 0 \\ 0 \\ 0 \end{pmatrix},~
v_L = \begin{pmatrix} 0 \\ 0 \\ 0 \\ -\sqrt{2E} \end{pmatrix},
\eea
where $E$ is the energy (or momentum) of the massless lepton.
In these expressions, $u(p,\lambda)$ have been defined for $p$ in the 
$\mathbf{\hat{z}}$ direction and $v(p,\lambda)$ have been 
defined for $p$ in the $-\mathbf{\hat{z}}$ direction.
Our conventions here are those 
in~\cite{Hagiwara:1986vm,Hagiwara:1985yu}; these
were chosen so that our $q\bar{q} \to ZZ^{(\ast)}$ helicity amplitudes
would reduce to those in Ref.~\cite{Hagiwara:1986vm} in the limit
where the $Z$ bosons are on shell.  

\subsection{Production amplitudes for the signal}

Because the Higgs is a scalar particle, 
the two $Z$ bosons can only have the following three helicity combinations: 
$(0,0)$ and $(\pm 1, \pm 1)$.  
Since the gluon-gluon-Higgs coupling is given by
\begin{equation}
  \frac{\alpha_s}{12\pi v} hG_{\mu\nu} G^{\mu\nu} \ ,
\end{equation}
where $v = 246$ GeV is the Higgs vev, 
the production helicity amplitude ${\cal M}^{ZZ}_{h;\lambda_1\lambda_2}$ 
for $gg\to h\to ZZ^{(\ast)}$ can be written as
\bea\label{h amp}
{\cal M}^{ZZ}_{h;\pm 1 \pm 1} &=& \frac{\alpha_s m_Z^2 \hat{s}} {3\pi v^2((\hat{s}-m_h^2)^2 + m_h^2 \Gamma_h^2)^{1/2}} \ , \\
{\cal M}^{ZZ}_{h;00} &=& \gamma_1 \gamma_2 (1 + \beta_1 \beta_2)\frac{\alpha_s m_Z^2 \hat{s}} {3\pi v^2((\hat{s}-m_h^2)^2 + m_h^2 \Gamma_h^2)^{1/2}}  \ .
\eea
This is the amplitude for a particular spin and color configuration
for the initial gluons.  In the interest of clarity, we do not
write the gluon helicities in the above amplitude, however these amplitudes
should be taken as the amplitude for the $++$ or $--$ initial state gluon 
helicities.  For the other two helicity combinations, the amplitude vanishes.
We will average the squared sum of these amplitudes
over spin and color when finding the differential cross section. 
It is worth pointing out that, as one can easily see from
Eq.~(\ref{h amp}),
in the high energy limit when the Higgs is heavy, 
the boost factor $\gamma \gg 1$ and the amplitude for two 
longitudinal $Z$ bosons, $(\lambda_1,\lambda_2)=(0,0)$, 
dominate over those for the transverse $Z$s.

\subsection{Production amplitudes for the background}

The production helicity amplitude for $q\bar{q}\to Z_1Z_2$ 
in the CM frame reads~\cite{Hagiwara:1986vm}
\be\label{back amp}
 {\cal M}^{ZZ}_{\sigma\bar{\sigma};\lambda_1\lambda_2}= 4\sqrt{2} \, 
\left(g_{\Delta{\sigma}}^{Zq\bar{q}}\right)^2 \, \epsilon\, 
\delta_{|\Delta\sigma|, \,\pm 1} 
\frac{{\cal A}^{\Delta\sigma}_{\lambda_1\lambda_2}(\Theta)\ 
d_{\Delta\sigma, \Delta\lambda}^{J_0}(\Theta)}{4\beta_1\beta_2\sin^2\Theta+
(1-\beta_1\beta_2)^2-x^2(1+\beta_1\beta_2)^2} \ .
 \ee
In the above $\Delta\sigma=\sigma-\bar{\sigma}$, 
$\epsilon=\Delta\sigma(-1)^{\lambda_2}$, 
$\Delta\lambda=\lambda_1-\lambda_2$,
and $J_0={\rm max}(|\Delta\sigma|,|\Delta\lambda|)$.
Note that in the limit of massless quarks, which we consider in this
work, the amplitude in Eq.~(\ref{back amp}) vanishes for $\Delta \sigma
= 0$.
The left- and right-handed coupling of quarks to the $Z$ boson
are given by:
\be\label{g}
g_{\Delta{\sigma}}^{Zq\bar{q}} = \frac{|e|(T_3 - 
Q \sin^2{\theta_W})}{\sin{\theta_W}\cos{\theta_W}},
\ee
where $T_3 = 0$ for right-handed quarks.
 Furthermore, $d_{\Delta\sigma, \Delta\lambda}^{J_0}(\Theta)$ is the 
$d$ function 
in the convention of  the Particle Data Group~\cite{Nakamura:2010zzi}. 
The coefficients ${\cal A}^{\Delta\sigma}_{\lambda_1\lambda_2}$ are
\bea
\Delta\lambda=\pm2&:& \  {\cal A}^{\Delta\sigma}_{\pm\mp} = 
-\sqrt{2}(1+\beta_1\beta_2) \ , \\
\Delta\lambda=\pm1&:& \ {\cal A}^{\Delta\sigma}_{\pm 0} = 
\frac{1}{\gamma_2 (1+x)} 
\bigg[(\Delta\sigma \Delta \lambda)
\bigg(1+\frac{\beta_1^2+\beta_2^2}{2}\bigg)-2\cos{\Theta}  \nonumber\\  
&& 
    -(\Delta \sigma \Delta \lambda) (\beta_2^2 - \beta_1^2)x -
    2 x \cos{\Theta} - (\Delta\sigma \Delta \lambda) 
    \bigg( 1 - \frac{\beta_1^2 + \beta_2^2}{2}\bigg)x^2\bigg] \, \\
&:& \ {\cal A}^{\Delta\sigma}_{0\pm} = 
\frac{1}{\gamma_1 (1-x)} 
\bigg[(\Delta\sigma \Delta \lambda)
\bigg(1+\frac{\beta_1^2+\beta_2^2}{2}\bigg)-2\cos{\Theta}  \nonumber\\  
&& 
    -(\Delta \sigma \Delta \lambda) (\beta_2^2 - \beta_1^2)x +
    2 x \cos{\Theta} - (\Delta\sigma \Delta \lambda) 
    \bigg( 1 - \frac{\beta_1^2 + \beta_2^2}{2}\bigg)x^2\bigg] \, \\
  \Delta\lambda=0&:& \  \ {\cal A}^{\Delta\sigma}_{\pm\pm} = -
  (1-\beta_1\beta_2) \cos\Theta
  - \lambda_1\Delta\sigma  (1+\beta_1\beta_2)x   \ , \\
  \Delta\lambda=0&:& \  \ {\cal A}^{\Delta\sigma}_{00} = 2\gamma_1\gamma_2 
  \cos\Theta \bigg[((1-x)\beta_1+(1+x)\beta_2)
    \sqrt{\frac{\beta_1\beta_2}{1-x^2}}-(1+\beta_1^2\beta_2^2) \bigg] \, .
  \eea
It can be checked that in the on-shell limit when $x\to 0$,
$\beta_1, \beta_2 \to \beta$, and $\gamma_1,\gamma_2 \to \gamma$, 
the above expressions reduce to those presented in 
Appendix D of Ref.~\cite{Hagiwara:1986vm}. 
Moreover, in the high energy limit when $\gamma\gg 1$, 
the $\Delta\lambda=\pm2$ amplitudes dominate, 
corresponding to $(\lambda_1,\lambda_2)=(\pm,\mp)$.

\subsection{Amplitudes for $Z$ boson decay to leptons}

The decay helicity amplitude for the $1\to 2$ process such as 
$Z\to \ell \bar{\ell}$ is essentially a matrix element of the 
spin-1 rotation matrix,  which is the inner product between states 
with definite projections of angular momenta along a chosen 
axis~\cite{Peskin:1995ev}. 
We have defined our angles, in particular $\theta_1$, $\theta_2$,
$\phi_1$, and $\phi_2$, such that we need consider only
$Z_i(\lambda) \to \ell_\sigma(\theta_i, \phi_i) 
\bar{\ell}_{\bar{\sigma}}$.  That is, the amplitude
has the same form for $i=1$ and for $i=2$.  Specifically, following the
conventions described above in 
Eq.~(\ref{definitions})~and~(\ref{spinor definitions}), we find that 
\begin{equation}\label{Zee}
{\cal M}^{(i)}_{\lambda_i;\sigma_i\bar{\sigma}_i} 
= \Delta \sigma_i~(-1)^{\lambda_i} 
\sqrt{2}~g_{\Delta{\sigma}}^{Z\ell\bar{\ell}}~
d(\Delta \sigma_i,\lambda_i,\theta_i)~m_i~e^{i \lambda_i \phi_i},
\end{equation}
where $d(\Delta \sigma_i,\lambda_i,\theta_i) = 
d_{\Delta \sigma_i,~\lambda_i}^{\text{max}(~|\Delta \sigma_i|,~|\lambda_i|~)}
(\theta_i)$ in the conventions of Ref.~\cite{Nakamura:2010zzi}.
In the above, $m_i$ is the invariant mass of $Z_i$, $\lambda_i$ 
is its helicity, $\Delta\sigma_{i}=\sigma_i-\bar{\sigma}_i$, 
and the coupling of the $Z$ to the lepton pair is
\be
g_{\Delta{\sigma}}^{Z\ell\bar{\ell}} =  
\frac{|e|(T_3 + \sin^2{\theta_W})}{\sin{\theta_W}\cos{\theta_W}},
\ee
where $T_3 = -1/2~~(0)$ for left (right)-handed leptons.
Note that the amplitude in Eq.~(\ref{Zee}) vanishes for $\Delta \sigma_i = 0$
in the limit of massless leptons, which is the limit taken in this work.
For $\Delta \sigma_i = -1$, the expression
in Eq.~(\ref{Zee}) reproduces Eq.~(4.8a) in 
Ref.~\cite{Hagiwara:1986vm}.

\subsection{Differential cross sections for signal and background}

The full amplitude of $q\bar{q}\to Z_1Z_2\to (\ell_1\bar{\ell}_1)
(\ell_2\bar{\ell}_2)$ is then
\be
{\cal M}(\sigma,\bar{\sigma};\sigma_i) = 
\sum_{\lambda_1,\lambda_2}  {\cal M}^{ZZ}_{\sigma\bar{\sigma};
\lambda_1\lambda_2}\ D_Z(k_1^2) D_Z(k_2^2)\ 
{\cal M}^{(1)}_{\lambda_1;\sigma_1\bar{\sigma_1}} 
{\cal M}^{(2)}_{\lambda_2;\sigma_2\bar{\sigma_2}} \ ,
\ee
where the propagator factor for the $Z$ boson may be taken to be
\be
D_Z(q^2)= \frac{1}{q^2-m_Z^2+ i m_Z \Gamma_Z}.
\ee
The full differential cross section is then (using the 
short-hand notation $\Omega=\{\Theta,\theta_1,\theta_2,\phi_1,\phi_2\}$ 
and $d\Omega=d\cos\Theta\, d\cos\theta_1\, d\cos\theta_2\, d\phi_1\, d\phi_2$)
\be\label{back diff}
\frac{d\sigma}{d\Omega\, dm_1^2 \, dm_2^2}= 2\pi \, \frac14\, \frac13\,  
\frac1{2\hat{s}}\, \frac{\beta_1(1+ x)}{32\pi^2}\left(\frac1{32\pi^2}\right)^2
\left(\frac1{2\pi}\right)^2 \ \sum_{\sigma,\bar{\sigma},\sigma_i} 
\left|{\cal M}(\sigma,\bar{\sigma};\sigma_i)\right|^2 \ ,
\ee 
where $2\pi$ comes from integrating over the unobservable production 
azimuthal angle, $1/4$ from averaging over initial spin states, 
$1/3$ from averaging initial color states, $1/(2\hat{s})$ from the incoming 
flux, $\beta_1(1+ x)/(32\pi^2)$ from the $ZZ^{(\ast)}$ two-body phase space, 
$1/(32\pi^2)$ from the phase space of each of the $\ell\bar{\ell}$ 
final states, and $1/(2\pi)$ from each $dm_i^2$ integral that one 
obtains when re-expressing the four particle phase space in terms of 
the $dm_i^2$ variables (see for example~\cite{muruyama phase space}).

Likewise, in the case of the 
$gg\to h \to Z_1Z_2\to (\ell_1\bar{\ell}_1)
(\ell_2\bar{\ell}_2)$ signal, the full amplitude is
\be\label{signal amp}
{\cal M}(h;\sigma_i) = 
\sum_{\lambda_1,\lambda_2}  {\cal M}^{ZZ}_{h;
\lambda_1\lambda_2}\ D_Z(k_1^2) D_Z(k_2^2)\ 
{\cal M}^{(1)}_{\lambda_1;\sigma_1\bar{\sigma_1}} 
{\cal M}^{(2)}_{\lambda_2;\sigma_2\bar{\sigma_2}} \ ,
\ee
and the full differential cross section is 
\be\label{signal diff}
\frac{d\sigma}{d\Omega\, dm_1^2 \, dm_2^2}= 2\pi \, \frac14\, \frac18\,  
\frac1{2\hat{s}}\, \frac{\beta_1(1+ x)}{32\pi^2}\left(\frac1{32\pi^2}\right)^2
\left(\frac1{2\pi}\right)^2 \ \bigg( 2 \bigg) \sum_{\sigma_i} 
\left|{\cal M}(h;\sigma_i)\right|^2 \ .
\ee 
Note that the only significant difference between Eq.~(\ref{back diff}) and 
Eq.~(\ref{signal diff}) is that color averaging yields a factor
of $\frac{1}{3}$ in Eq.~(\ref{back diff}), but $\frac{1}{8}$ in
Eq.~(\ref{signal diff}).  The factor of $2$ before the sum in 
Eq.~(\ref{signal diff}) is from summing over initial gluon helicities.

Before proceeding, we note our definitions of $Z_1$ and $Z_2$.
When the two $Z$ bosons decay into $2e2\mu$ final states, 
we define $Z_1$ to be the parent particle of the electron/positron pair,
while for $4e$ or $4\mu$ final states, $Z_1$ is the $Z$ with lower
invariant mass.  Technically, one could reconstruct e.g. a $4\mu$
final state in two ways (in terms of assigning a particular muon and 
a particular antimuon to a given $Z$), however generally 
one of these reconstructions would make the resulting $Z$s far off-shell.
Hence we only keep the reconstruction for which the product of 
the $Z$ Breit-Wigner factors is largest.
We note that the total event rate of $ZZ^{(\ast)} \to 2e2\mu$ is a factor of 
two larger than that of $ZZ^{(\ast)} \to 4e$ or $ZZ^{(\ast)}\to 4\mu$.

\subsection{Angular distributions}

Analytic expressions for the background and signal distributions are 
contained in Eqs.~(\ref{back diff}) and (\ref{signal diff}). However, 
they are too long to be presented explicitly  and not very illuminating. 
Instead we will show figures for all the singly and doubly distributions 
for the signal and the background in this subsection.

\begin{figure}
\includegraphics[scale=0.57, angle=0]{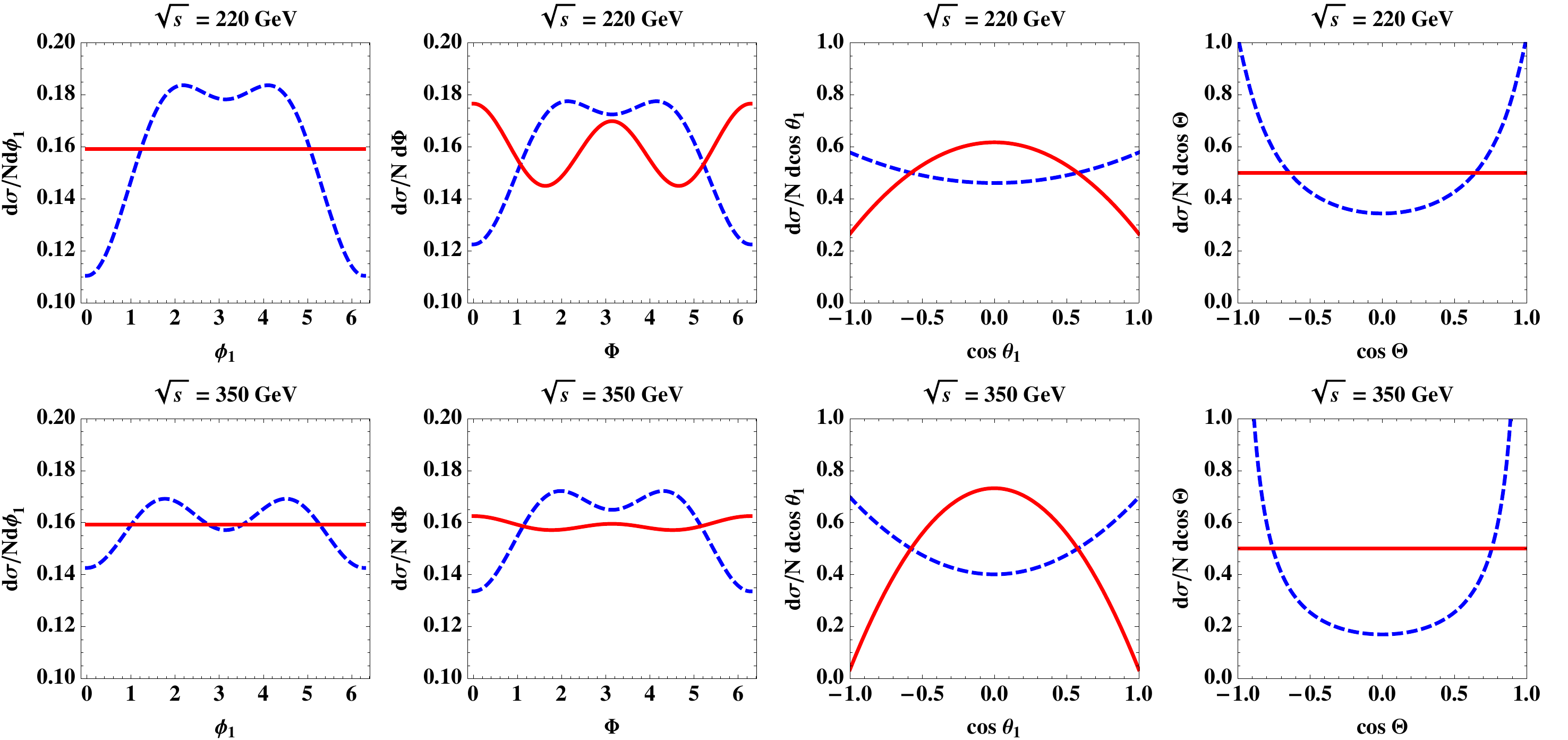}  
\caption{\label{sindist}{\em Signal and background singly differential 
distributions at $m_h = \sqrt{{s}}= 220$ and $350$ GeV. The blue (dashed) 
lines are background distributions and the red (solid) lines are signal 
distributions.
}}
\end{figure}

\begin{figure}
\includegraphics[scale=0.45, angle=0]{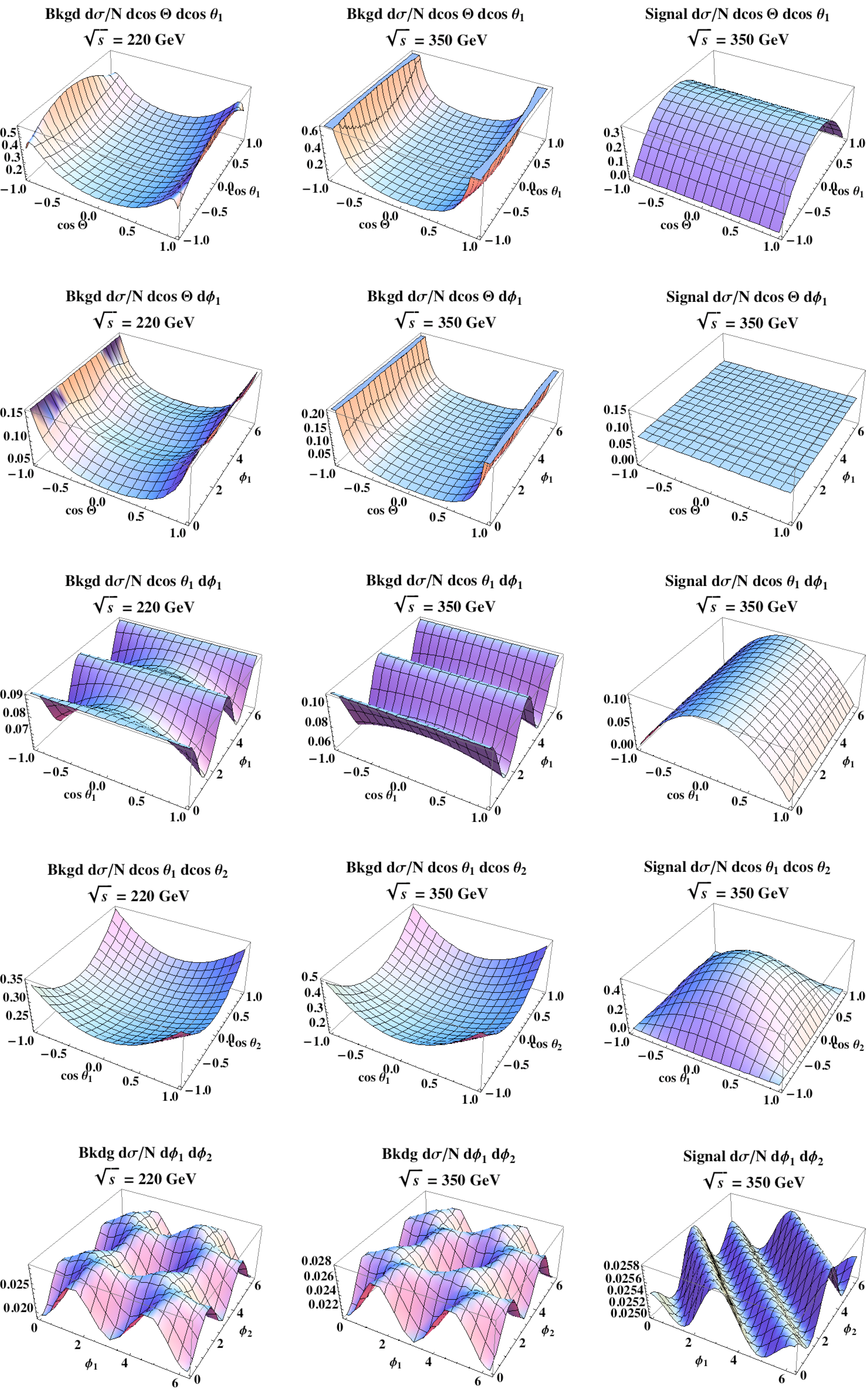}  
\caption{\label{doubdist}{\em Doubly differential distributions for 
signal and background. 
See the text for more explanations.
}}
\end{figure}

In Fig.~\ref{sindist} we show normalized singly differential 
distributions in $\phi_1, \Phi\equiv \pi - \phi_1 - \phi_2, \Theta,$ 
and $\theta_1$ for the processes 
$u\bar{u}\to ZZ\to 4\ell$ and $h\to ZZ\to 4\ell$ at $m_h = \sqrt{{s}}= 220$ 
and 350 GeV. (For simplicity, we take the $Z$ bosons to be on-shell.)
One notable feature is that the signal $\phi_1$ distribution is 
flat, although there are correlations between $\phi_1$ and $\phi_2$ such that 
$\Phi$ exhibits a $\cos 2\Phi$ dependence. It was pointed out in 
Refs.~\cite{Keung:2008ve,Cao:2009ah} that this observable is very useful 
in discerning the spin and CP properties of a singly produced resonance 
decaying to two $Z$ bosons.
In addition, distributions in the production angle $\Theta$ indicate 
that the $ZZ$ pair produced by the background process tend to be in the 
forward region inside the detector; 
this is especially pronounced when the invariant mass is 
high. This is a feature of the $t$-channel production mechanism which has 
been suggested as a way to tag the vector boson fusion production of the 
Higgs boson, where there are two forward jets in the event. In the signal
case, on the contrary, the $Z$s are produced isotropically, 
as expected from the fact that the Higgs is a scalar particle.

It is also interesting to consider doubly differential angular 
distributions for signal and background, which are shown in 
Fig.~\ref{doubdist}. We plot distributions in the following pairs of 
angles: $(\Theta, \theta_1)$, $(\Theta, \phi_1)$, 
$(\theta_1, \phi_1)$, $(\theta_1, \theta_2)$, and $(\phi_1, \phi_2)$. 
The background distributions are again from $u\bar{u}$ initial states and 
shown for $\sqrt{s}=220$ and $350$ GeV. We observe noticeable changes in 
the distributions of the first three pairs of variables for these two 
different center-of-mass energies. On the other hand, the signal 
distributions do not change as much when one varies the Higgs mass, 
except for the $(\phi_1, \phi_2)$ distributions, and are shown only for 
$m_h=350$ GeV. The background exhibits strong correlations 
between the pairs of angles in all five cases.

\section{Statistical Procedures}\label{statistics}

Likelihood methods are frequently employed to establish the presence, or
lack, of a signal using kinematic distributions which
discriminate signal from background.  For example, if the purpose is discovery,
one considers the ``null hypothesis'' assuming the observed data set is 
entirely due to background and an ``alternative hypothesis'' assuming 
the presence of both signal and background.  
A likelihood function is defined for each hypothesis to quantify the 
probability of obtaining the actual data under that particular hypothesis. 
In order to accept or reject one hypothesis in favor of the other, a 
``test statistic'' must also be defined.

In quantum field theory there is a natural object quantifying the probability 
of obtaining a particular event in a given set of data: the 
differential cross section. 
This motivates the use of the MEM, which is simply the use of likelihood 
methods where the probability density function 
({pdf}) used in the likelihood is the properly normalized 
differential cross section (or ``matrix element'') 
with respect to certain kinematic 
variables\footnote{We use ``pdf'' for probability density function and ``PDF'' 
for parton distribution function.}.
When the number of events is small one can include every event 
in the evaluation of the likelihood. 
This  ``unbinned likelihood method'' is what we adopt in the following.

\subsection{Extended maximum likelihood method}\label{EML subsec}

When there exist free parameters in the underlying hypothesis one wishes 
to test\footnote{In such cases the hypothesis is called ``composite'',
 while those without free parameters are called ``simple''.}, statistically 
preferred values of the parameters of the underlying model are found by 
maximizing the likelihood function with respect to those parameters.  
When the overall number of events is not fixed but allowed 
to fluctuate, the normalization of the likelihood function may become a 
free parameter.  In this case, one is using the ``extended'' 
maximum likelihood method~\cite{Barlow:1990vc}, which we employ in this
work.

To be more specific, we consider the likelihood for some collider
signature with (unknown) expected number of events $\mu$ described
by kinematic information $\boldsymbol{x}=\{x_1, x_2, ... x_N\}$, where
the $x_i$ are kinematic variables describing event $i$ of $N$ total
events.  The unbinned likelihood is simply
\begin{equation}\label{EML 1}
  {\mathcal{L}}(\mu; \boldsymbol{\theta}) 
  = \frac{e^{-\mu}\mu^N}{N!} \prod_{i=1}^N P(\boldsymbol{\theta}; x_i),
\end{equation}
where $P(\boldsymbol{\theta}; x)$ is the pdf for the kinematic variables 
as a function of $\boldsymbol{\theta}$, the parameters of the underlying
model.  In the MEM, $P(\boldsymbol{\theta},x)$ is the differential 
cross section normalized by the total cross section, 
generally scaled by efficiencies and acceptances of the detector involved.

We will define  $P_s(m_h,x_i)$ to be the normalized pdf for the signal 
process, which depends on one underlying parameter, the Higgs mass 
$m_h$.\footnote{We set the Higgs width to the SM value
for the given Higgs mass, $m_h$, as found by HDECAY~\cite{Djouadi:1997yw}.}
The normalized pdf for the background process is then given by
$P_b(x_i)$; it is assumed here that the differential cross section 
for the background process is known and contains no unknown parameters.
The likelihood function for the signal plus background hypothesis is then 
given by
\begin{equation}\label{EML 4}
  {\mathcal{L}}_{s+b}(\mu,f,m_h) 
  = \frac{e^{-\mu}\mu^N}{N!} \prod_{i=1}^N 
  [ f P_s(m_h; x_i) + (1-f) P_b(x_i) ]\ ,
\end{equation}
where $\mu = \mu_s + \mu_b$ is the sum of the expected number of signal 
events $\mu_s$ and expected number of background events $\mu_b$, 
while the signal fractional yield $f$ is defined as
\begin{equation}\label{f}
0 \ < \   f=\frac{\mu_s}{\mu_s + \mu_b} \ < 1 \ .
\end{equation}
We often refer to $f$ simply as the ``yield''. For ${\cal L}_b$, 
the likelihood for the background-only hypothesis, we simply set $\mu_s=0$. 
When computing the likelihood for each hypothesis, we calculate the maximum
of ${\cal L}_{s+b}$ in $\{\mu, m_h, f\}$ and  ${\cal L}_b$ in $\mu$, 
respectively\footnote{In our fitting procedure, we look for 
the local maximum closest to the true Higgs mass.}.
In practice, since the Poisson distribution factors in the definition 
of the likelihood function, the maximum of the 
likelihood always occurs at $\mu=N$. 
So effectively one can replace $\mu$ by $N$, 
the measured total number of events, in the calculation.
 
\subsection{Expected significance}\label{significance subsection}

In determining the significance for Higgs discovery from a set of events, 
we are really comparing the likelihood of two hypotheses:
(i) that the events consist of signal events from the 
Higgs boson at some mass, as well as events from background
$q\bar{q} \to ZZ^{(\ast)}$ production, and
(ii) that all the events are due to $q\bar{q} \to ZZ^{(\ast)}$ production.
Our choice of test statistic for describing this relative likelihood is
the likelihood ratio
 \be
 {\mathcal Q} = \frac{{\cal L}_{s+b}}{{\cal L}_b} \ ,
 \ee
 from which the significance of discovery is computed
 \begin{equation}\label{significance}
  {\mathcal S} = \sqrt{2 \ln {\mathcal Q}} \ .
\end{equation}
This test statistic was used by the LEP experiments in their 
Higgs searches~\cite{Barate:2003sz}, while its use for
$h \to ZZ^{(\ast)} \to 4\mu$ was studied in Ref.~\cite{cmsQ}. 
In our case we include the angular correlations in the likelihood 
function. The expected significance ${\cal S}$ is then obtained by performing 
a large number of pseudo-experiments and choosing the median value. The 
$1$~and~$2$~$\sigma$ spreads in ${\cal S}$ are determined from the distribution 
of significances obtained from the pseudo-experiments.

\subsection{Exclusion limit}\label{yield limit}

We determine the exclusion limit, in the absence of a signal, by setting 
an upper limit on the yield, $f$, defined in Eq.~(\ref{f}). For a particular 
choice of Higgs mass $\hat{m}_h$, we define a pdf $f$ by 
considering the likelihood ${\cal L}_{s+b}$ as a function of $f$,
\begin{equation}\label{p(f)}
  p(f) = \frac{{\mathcal{L}_{s+b}}(N, f,\hat{m}_h)}
  {\int_0^1 {\mathcal{L}_{s+b}}(N, \bar{f},\hat{m}_h)~d\bar{f}} \ .
\end{equation}
The $95\%$ confidence level limit on $f$ for a given set of data is given 
by $\alpha$ as follows:
\begin{equation}
  \int_0^\alpha p(f)~df = 0.95.
\end{equation}
We then translate $\alpha$ into a $95\%$ confidence level upper limit on the 
Higgs production cross section by unfolding with the detector acceptances and 
efficiencies. The expected exclusion limit is obtained by performing a large 
number of pseudo-experiments.

It should be noted that the procedure above for setting the exclusion limit 
only takes into account differences in the shape of kinematic distributions 
between signal and background. In particular we are mainly interested in 
possible improvements by including angular distribution in addition to the 
invariant mass spectrum. Typically experimental collaborations set limits 
directly on the normalization of the signal cross section by performing 
counting experiments in a particular window of total invariant mass. 
In comparison with the
standard CL$_s$ method employed by ATLAS and CMS collaborations, our
method should be considered as a shortcut for the purpose of
understanding the improvement from incorporating the angular
correlations.
One would hope to incorporate both the counting experiments and shape 
measurements in a more complete study.

\subsection{Probability density functions}\label{sect:pdf}

In this subsection we define the signal and background pdfs that enter into 
the likelihood function in Eq.~(\ref{EML 4}). The kinematic observables are
$\boldsymbol{x}=\{Y, \hat{s}, m_1^2, m_2^2, \Omega\}$, where $Y$ is the 
pseudo-rapidity of the $ZZ^{(\ast)}$ system, $\hat{s}$ is the partonic 
center-of-mass energy, $m_{1(2)}$ is the invariant mass of the $Z_{1(2)}$ 
boson, and $\Omega$ represents the production and decay angles.

The signal pdf is
\begin{equation}
 {P}_s(m_h;\boldsymbol{x} ) = \frac{1}{\epsilon_s \sigma_s(m_h)}
\frac{d\sigma_s (m_h;  \boldsymbol{x})}
{dY\, d\hat{s}\, dm^2_1\, dm^2_2\, d\Omega},
\end{equation}
where $\sigma_s(m_h)$ is the total hadronic cross section,
$d\sigma_s$ is the corresponding differential cross section,
and $\epsilon_s$ is the total signal efficiency (which in principle
includes geometric acceptance as well as reconstruction efficiencies).
More explicitly,
\begin{equation}\label{sig prepdf} 
 {P}_s(m_h;\boldsymbol{x} ) =
\frac1{\epsilon_s \sigma_s(m_h)} \bigg( \frac{f_g(x_1) f_g(x_2)}{s} \bigg)
\frac{d\widehat{\sigma}_{h}(m_h,\hat{s}, m_1, m_2, \Omega)}
{dm^2_1\, dm^2_2\, d\Omega}\ .
\end{equation}
In the above, $\widehat{\sigma}_{h}$ is the partonic cross section,
and the $f_g(x)$ is the gluon PDF.

For the background pdf we take into account the fact that,
in a hadron collider, we are unable to determine the direction of 
the initial quark (as opposed to the anti-quark) on an event-by-event
basis by calculating the cross section for each
choice of initial quark direction and summing the two. This may be written as
\begin{equation}\label{back prepdf}
 {P}_b(\boldsymbol{x}) =  
 \frac1{\epsilon_b \sigma_{q\bar{q}} } 
 \bigg(\bigg( \frac{f_q(x_1) f_{\bar{q}}(x_2)}{s} \bigg)
 \frac{d\widehat{\sigma}_{q\bar{q}}(\hat{s}, m_1, m_2, \Omega)} 
 {dm^2_1\, dm^2_2\, d\Omega}\ + 
 \bigg( \frac{f_{\bar{q}}(x_1) f_q(x_2)}{s} \bigg)
 \frac{d\widehat{\sigma}_{q\bar{q}}(\hat{s}, m_1, m_2, \Omega^\prime)} 
 {dm^2_1\, dm^2_2\, d\Omega^\prime}\ \bigg),
\end{equation}
where $\Omega^\prime \equiv (\pi - \Theta, \theta_1, \theta_2, \phi_1 + \pi, \phi_2 + \pi)$ is the shift in angles needed for an initial quark in the $-z$ direction and we have switched the quark PDF with the anti-quark PDF (or equivalently switched $x_1$ and $x_2$). The total $q\bar{q} \to ZZ^{\ast} \to 4\ell$
cross section is given by $\sigma_{q\bar{q}}$ with the function $f_{q(\bar{q})}$ representing the quark (anti-quark) PDF,
and $\epsilon_b$ the total efficiency for this channel.
In calculating the cross section, $\widehat{\sigma}_{q\bar{q}}$, 
for the $q\bar{q} \rightarrow ZZ^{(\ast)} \rightarrow 4\ell$ background, 
we sum over the quark flavors, $u, d, s$ and $c$.
We use CTEQ5L for the PDFs 
indicated in Eq.~(\ref{sig prepdf})~and Eq.~(\ref{back prepdf}) 
as $f_i(x)$~\cite{Lai:1999wy}.

\section{Monte Carlo Simulations}\label{angular2}

We generate the signal and background events for our analysis using 
MadGraph/MadEvent (MG/ME) version 4.4.52 \cite{Alwall:2007st}. 
Proton-proton collisions 
at $\sqrt{s}=7$ TeV are implemented with the CTEQ5L~\cite{Lai:1999wy} 
PDFs. As noted above, 
the main irreducible background, and the only one which 
we will consider in our analysis, is 
$q\bar{q} \rightarrow ZZ^{(\ast)} \rightarrow 4\ell$.
For signal we consider only 
$gg \to h \rightarrow ZZ^{(\ast)} \rightarrow 4\ell$, 
without additional jets in the final states;
effectively our analysis considers only the ``0-jet bin'' for the 
$ZZ^{(\ast)} \to 4\ell$ channel.

\begin{figure}
\includegraphics[width = 6 in]{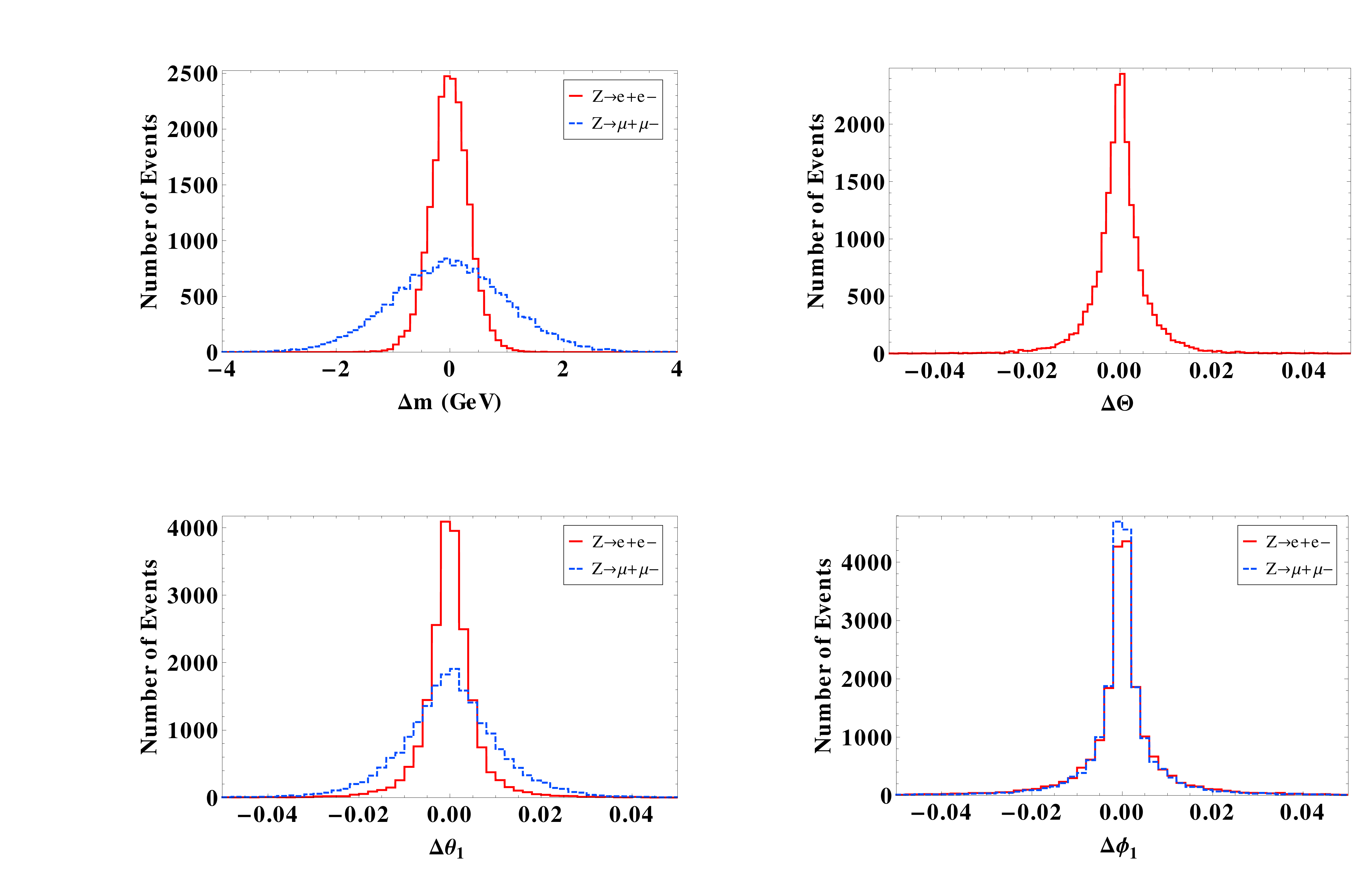} 
\caption
{\label{fig : smeareffects}\emph{Effect of detector resolution in 
measuring $E$ and $p_T$ on kinematic variables. 
Each of the histograms were generated using twenty thousand events. 
The channel used for these plots is the background 
$2e2\mu$ channel.
 }}
\end{figure}

Detector effects are modeled by applying Gaussian
smearing to the energy of electrons according to 
\begin{equation}\label{electron smearing}
\Big(\frac{\sigma_{E,e}}{E}\Big)^2  =    \Big(\frac{0.036}{\sqrt{E}} \Big)^2 
+ 0.0026^2,
\end{equation}
and to the $p_T$ of muons according to
\begin{equation}\label{muon smearing}
\sigma_{p_T,\mu}= 0.015~p_T - 5.7 10^{-6}~p_T^2 + 2.2 10^{-7}~p_T^3.
\end{equation}
These expressions follow the CMS TDR \cite{Bayatian:2006zz}.  
The value of the constant term in Eq.~(\ref{electron smearing}) may be 
somewhat optimistic. However, we verified that our results do not change 
significantly when using a value for this quantity that is twice as large.
The smearing of the muon $p_T$ 
is quite conservative as we have used the resolution 
corresponding to $1.8<|\eta|<2.0$ and assumed this to be the same for lower
values of $\eta$ as well. The angles measured in the lab frame are not 
smeared. Nevertheless, angles
which define the kinematics of the four lepton system,  as described in 
Sect.~\ref{kinematics},  are affected by the $E$ and $p_T$ resolution,
as can be seen in Fig.~\ref{fig : smeareffects}.
A more sophisticated treatment of detector effects would include the effects
of reconstruction efficiencies, which we neglect for simplicity.

Another effect of smearing the lepton $E$ or $p_T$ is that 
the four lepton system  will now have a small $p_T$ in  the lab frame, 
even without the presence of additional jets or particles in the final 
states.  We simply boost away this induced $p_T$ 
and proceed to define angles as in Section~\ref{kinematics}.

\begin{table}[t]
\begin{center}
\begin{tabular}{|c|c|c|c|c|}
\hline
\multirow{7}{*}{Signal } & $m_h $(GeV) & ~$\sigma$(fb)~&~$\epsilon$~
&~$\langle N \rangle$ \\
\cline{2-5}
 & 175  & 0.218  &  0.512 & 0.279  \\
 & 200 &  1.26 &  0.594  & 1.87 \\
 &220 &   1.16  &  0.625 & 1.81  \\
 &250 &   0.958 &  0.654 &1.57\\
 &300 &   0.714 &   0.701 & 1.25\\
 &350 &   0.600 &   0.708 &  1.06\\
 \hline
 Background & - & 8.78 & 0.519 & 11.4 \\
\hline
\end{tabular}
\caption{\label{tab:DM_prof}
\emph{Expected number of events $\langle N \rangle$ 
for the signal and background $2e2\mu$ channel at $2.5$fb$^{-1}$. 
The SM cross section without any cuts is denoted by $\sigma$, 
while $\epsilon$ is the efficiency of our analysis cuts. 
The 4e and $4\mu$ channel yields are half of the $2e2\mu$ 
channel and their efficiencies are the same as the $2e2\mu$ channel.}}
\end{center}
\end{table}

After generating events and smearing the lepton $p_T$, we apply the following 
cuts to each of the lepton in the lab frame:
\bea
|p_T| &\ge& 10\quad {\rm GeV}~, \nonumber \\
|\eta| &\le& 2.5~. \nonumber 
\eea
Moreover, we focus on the following window for  the $4\ell$ invariant mass
\be
150 \quad {\rm GeV} \ \le\  \hat{s} \  \le\ 450 \quad {\rm GeV}  \ .
 \ee
\begin{figure}
\includegraphics[width = 6 in ]{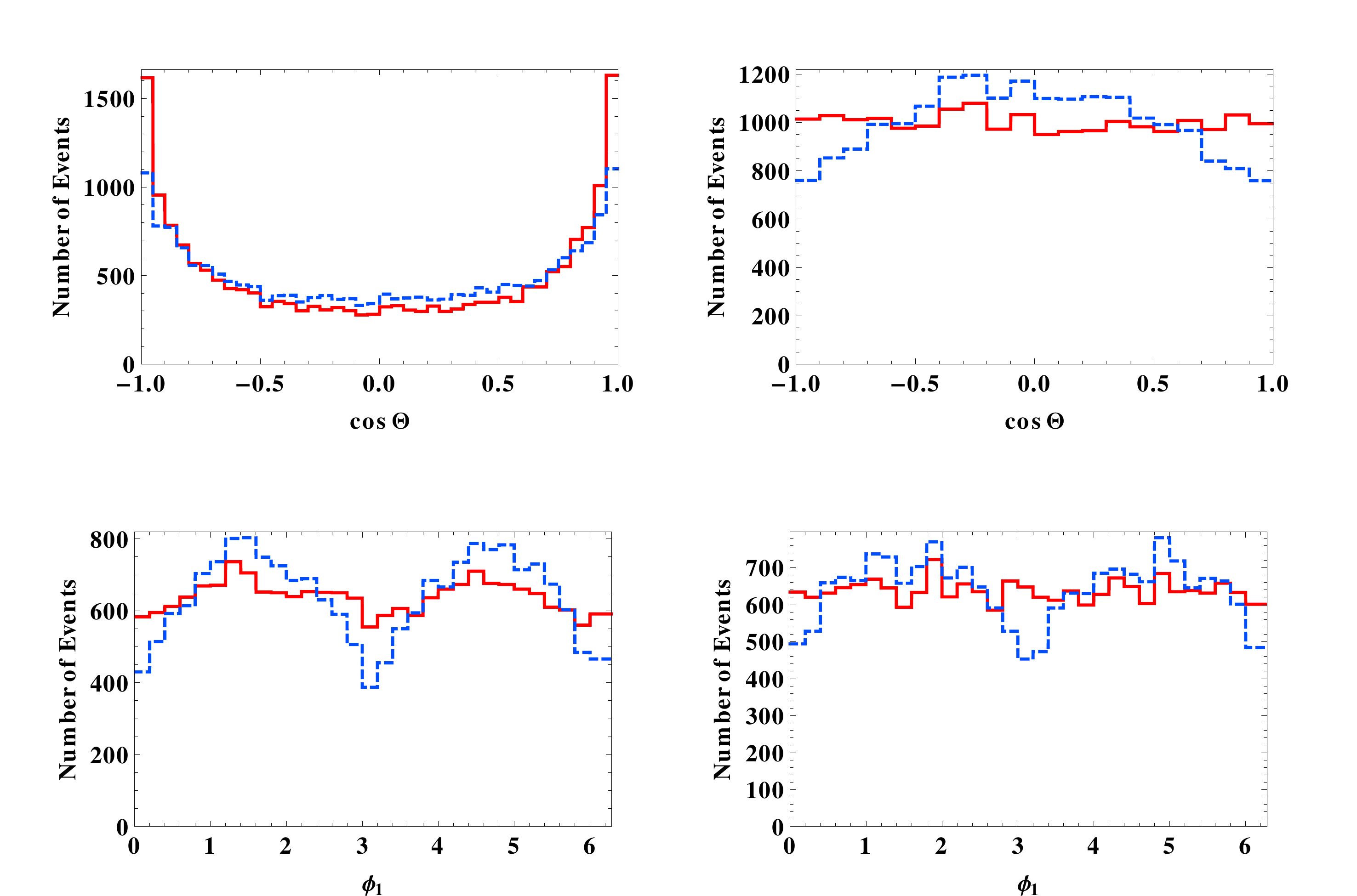} 
\caption{\label{fig:accdists}\emph{The effect on $\cos{\Theta}$ and $\phi_1$ 
distributions from rapidity and transverse momentum cuts on final state 
leptons. The panels on the left show the distribution for background events 
before (red) and after (blue) the cuts. The panels on the right show the
distributions for signal events before (blue) and after (red) the cuts. 
Signal distributions correspond to a SM Higgs boson with $m_h=300$ GeV.}} 
\end{figure}
The efficiency (geometric acceptance) for our selection of signal events is 
listed in Table~\ref{tab:DM_prof} and varies from $\sim 0.5$ to $\sim 0.7$ 
as we increase the mass of the Higgs boson from $170$ GeV to $350$ GeV. 
The efficiency for selection of background events is $0.52$. Moreover, 
these cuts affect the reconstructed angular distributions.
As can be seen in Fig.~\ref{fig:accdists}, the distributions with respect 
to  $\Theta$, $\phi_1$ and $\phi_2$ are modified most significantly by our 
acceptance cuts. In the case of the $\Theta$ distribution it is easy to see 
that the loss of events in the forward region due to $p_T$ and rapidity cuts 
cause a change in the shape of the distribution. 
The $\phi_1$ and $\phi_2$ distributions undergo less trivial modifications.

Also listed in Table \ref{tab:DM_prof} are the cross sections and expected 
number of events after selection cuts, with an integrated luminosity of 
$2.5$~fb$^{-1}$,  for both signal and background at the LHC with 
$\sqrt{s}=7$~TeV. 
The cross sections for the signal and background processes are obtained 
from MG/ME with $K$-factors applied.
As we are considering the $0$-jet bin for signal and background, there
are a number of complications in determining the 
appropriate higher order cross section~\cite{Catani:2001cr},
a complete study of which is beyond the scope of this paper.  
Therefore we simply apply a $K$-factor of $1.5$ for the signal 
and a $K$-factor of $1.33$ for the background.  
(The $K$-factor for the background is the value obtained in 
Ref.~\cite{Barger:1988fm}.)  We note that these $K$-factors,
are not used in the analyses themselves, 
but only in determining the number of events of each type to 
include in the pseudo-experiments.

\section{Results}\label{results}

In this section we present our results for the
expected significance and expected exclusion limits from the MEM, 
which includes the following kinematic variables
for the four lepton final state, 
\be
\boldsymbol{x} = \{Y, \hat{s}, m_1^2, m_2^2, \Theta, \theta_1, \theta_2, 
\phi_1, \phi_2\} \ .
\ee 
For comparison we also present results from an analogous analysis,
which uses only invariant mass information in the pdf.  
(Specifically we consider the product of Breit-Wigner factors for 
the Higgs as our signal pdf and use the invariant mass
distribution from simulated data as the background pdf.)
Our results are for the LHC at $\sqrt{s}=7$ TeV and integrated 
luminosities of $2.5$, $5$, and $7.5$~fb$^{-1}$  for discovery 
and $1$, $2.5$, and $5$~fb$^{-1}$ for exclusion.

\subsection{Expected significance}

To compute the expected significance, we perform ten thousand 
pseudo-experiments (each) for Higgs masses of $175$, $200$, $220$, 
$250$, $300$, and $350$ GeV. 
Each pseudo-experiment consists of signal and background events with 
$4\mu$, $2e 2\mu$, and $4e$ final states; the  number of events
of each type are chosen from Poisson distributions, 
where the expected numbers of signal and background events 
are given by the product of the luminosity under consideration, 
the theoretical cross section, and acceptance efficiencies 
after smearing and cuts.  
The cross section after cuts is also used to normalize the signal and 
background pdfs $P_s$ and $P_b$ in the likelihood function in 
Eq.~(\ref{EML 4}).
As we do not include the different reconstruction efficiencies
for electrons and muons in our analysis,  $P_s$ and $P_b$ are identical
for $4\mu$, $2e 2\mu$, and $4e$ final states. However, we do
consider the yields in each channel as separate parameters when
finding the significance from pseudo-experiments
by maximizing the likelihood  with respect to the undetermined parameters.

\begin{figure}

\includegraphics[width = 4 in]{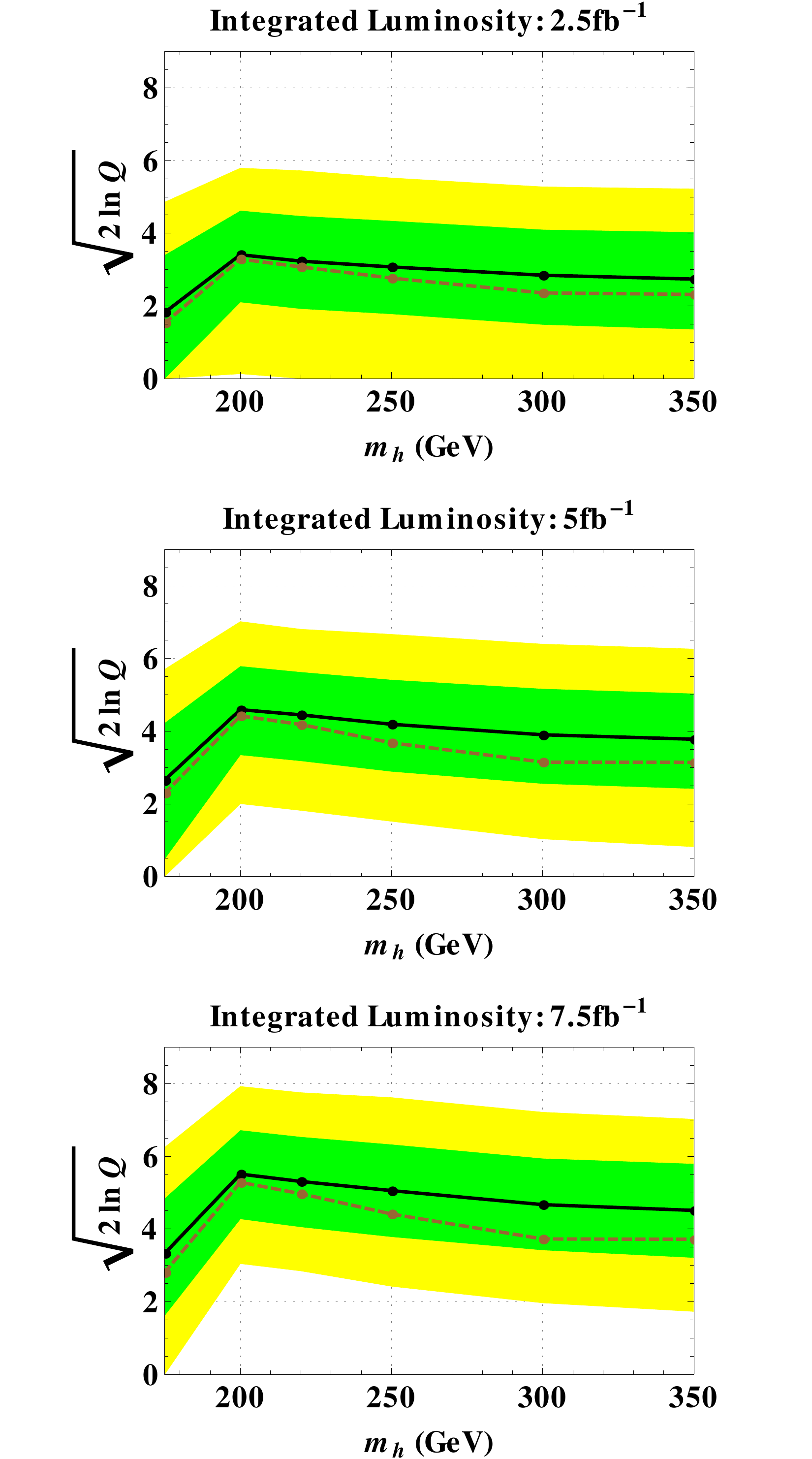} 

\caption{\label{fig : sig} 
\emph{Comparison of significance for Higgs discovery obtained from using 
only invariant mass information (brown dotted line) with that obtained  
from all of the kinematic variables using the MEM (black solid line) at 
different integrated luminosities. The green and yellow bands correspond 
to less than 1 and 2 $\sigma$ deviations from the median expected 
significance in the case where we use all kinematic variables. }}
\end{figure}

The median values of significance obtained for each Higgs mass
in this channel, using both the MEM (all kinematic variables) and the 
likelihood method (invariant mass only),
are shown in Fig.~\ref{fig : sig}. 
The $1\sigma$ and $2\sigma$ bands on the significance 
obtained from the MEM are also shown in this figure.
We do not show the corresponding bands for the invariant
mass-based analysis; the widths of the bands 
for this case are similar.

We note that the MEM consistently outperforms the invariant 
mass only method, and that the effect is more pronounced, 
on the order of 10 - 20 \%, at higher Higgs masses.  
This is because the helicity amplitudes for $h\to ZZ^{(\ast)}$ and 
$q\bar{q} \to ZZ^{(\ast)}$ 
are increasingly different at larger values of invariant mass.  
More specifically, as already pointed out in Sect.~\ref{diffx},  the 
$(\lambda_1,\lambda_2) = (\pm 1, \mp 1)$ amplitudes dominate in the high
mass case for $q\bar{q}$, while only 
the $(\lambda_1,\lambda_2) = (0, 0)$ amplitude survives 
in the heavy Higgs limit in the $h\to ZZ^{(\ast)}$ case. 

\subsection{Exclusion}

\begin{figure}

\includegraphics[width=6.5 in]{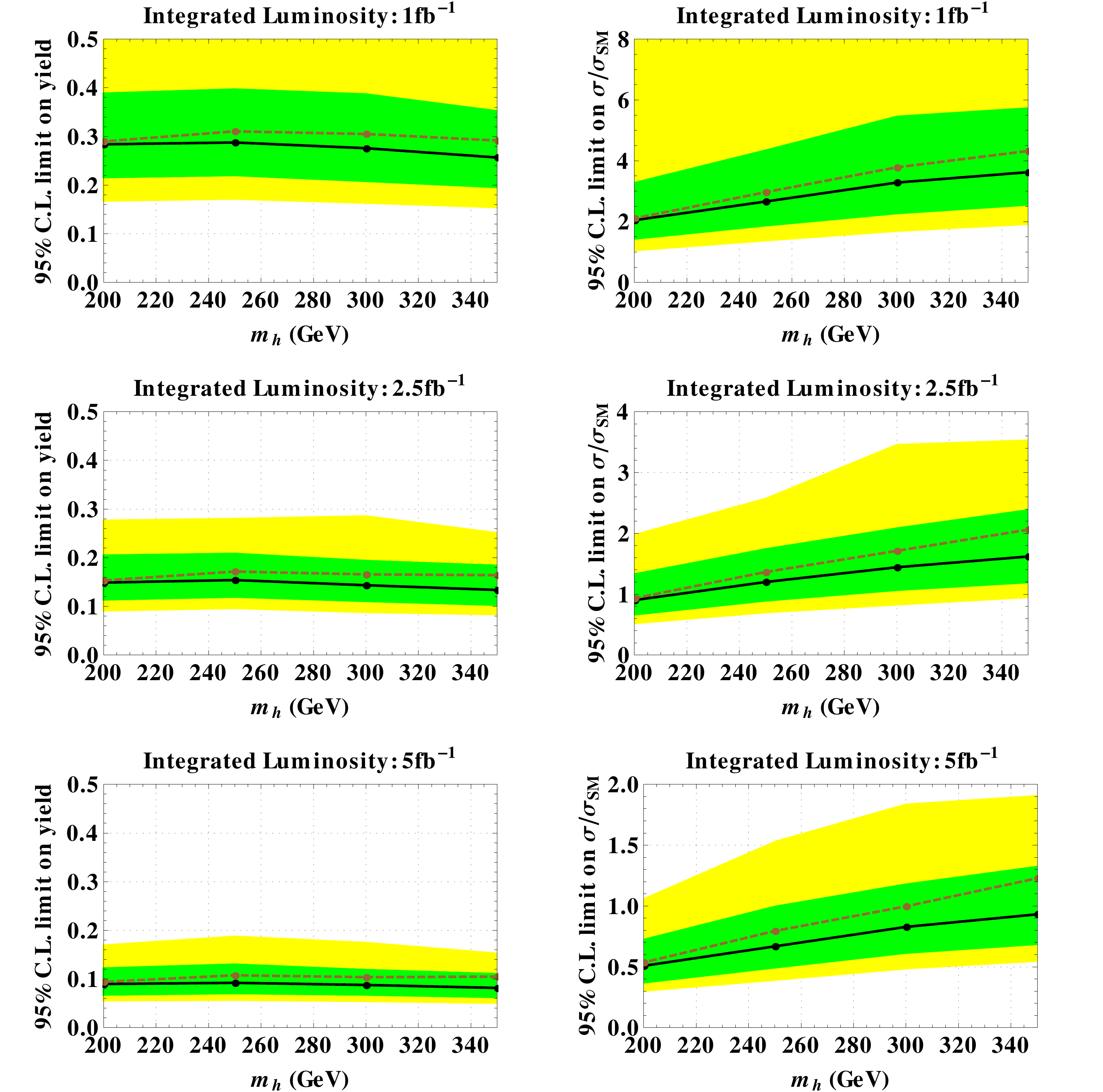}

\caption{\label{fig : exc}
\emph{The panels on the left contain $95\%$ confidence level exclusion limits
on the Higgs signal yield, $f$, while the panels on the right contain
these limits interpreted as limits on the Standard Model Higgs cross 
section.  The exclusions represented by the  brown, dashed line were 
obtained using only the invariant mass distribution, 
while all of the kinematic variables were used (via the MEM) 
in obtaining those exclusions represented by the black solid line. 
The green and yellow 
bands correspond to less than $1$~and~$2$~$\sigma$ deviations from the 
median expected exclusion in the case where we use all kinematic variables.
These results were obtained by performing one thousand pseudo-experiments.}}
\end{figure}

In Figure~\ref{fig : exc},
we show the median $95\%$ confidence level exclusions on the yield parameter 
$f$, both as obtained with the MEM and as obtained with a likelihood method 
using only invariant mass information. 
These
limits were obtained using the procedure described in 
Subsection~\ref{yield limit} to find the $95\%$ confidence limit on
$f$ in a given pseudo-experiment consisting only of background 
events, which were generated, as in our investigation of significance, 
using the method described in Section~\ref{angular2}.
We note that in obtaining these limits, the yield, $f$, was fixed
to be the same for each of the three channels ($4e$, $2e2\mu$, and $4\mu$),
and one thousand pseudo-experiments were performed.

We can translate a limit on the yield to a limit on the Higgs production
cross section, since the yield is defined as the ratio of the expected
number of signal events to the expected number of signal and background
events.  As this is the expected number of events after the selection cuts 
are applied, it is proportional to the cross section times efficiency,
rather than simply the cross section.  Specifically, we find
\begin{equation}\label{sigma from f}
   \sigma_s = \bigg(\frac{\epsilon_b}{\epsilon_s}\bigg)
        \bigg(\frac{f}{1-f}\bigg) \sigma_b,
\end{equation}
where $\epsilon_{s(b)}$ and $\sigma_{s(b)}$ are the signal (background)
efficiency and cross section respectively.  
Eq.~(\ref{sigma from f}) allows us to translate our limits on $f$ into
a limit on $\sigma_s$, the Higgs cross section, as is done in
Fig.~\ref{fig : exc}.  For simplicity we do not include any systematic 
uncertainties (e.g. on the efficiencies) in the analysis.
We see that, as in the investigation of significance considered above, 
the difference in sensitivity between the MEM and the 
invariant-mass-only analysis is greater in the higher Higgs mass range.  

\section{Conclusions}\label{conclusions}

Traditional search strategies for the Higgs boson in the $ZZ^{(*)}\to
4\ell$ channel concentrate only on measurements of the total
invariant mass of the four leptons. Since the four-momenta of all final
state particles can be reconstructed, we have considered in this work
the possibility of including all available kinematic information in
discriminating the Higgs signal from the dominant irreducible
background at the LHC with $\sqrt{s}=7$ TeV, by implementing the
Matrix Element Method.

We first derived Lorentz-invariant expressions for the production and
decay angles used in the analyses, which allow for reconstruction of
these angular variables from momenta measured in the laboratory frame.
Then we presented analytic expressions for the fully differential
distributions for both signal and background, allowing the $Z$
bosons to be off-shell.

We found that greater sensitivity
in discovering or excluding the Standard Model Higgs boson can be
achieved when including spin correlations in addition to total
invariant mass measurements. Generally, these improvements are
on the order of $10$ - $20$ \%; they are larger for higher Higgs
masses, for which the differences between signal and
 background distributions
are greater.

Searching for the Higgs boson is among the top priorities of the physics
program at the LHC. Our results indicate it would be worthwhile to
include full angular distributions in actual experimental searches  in
the golden channel through some type of multivariate analyses. Whether
a Standard Model Higgs boson will be discovered or excluded in the
near future, we eagerly await the result!

\begin {acknowledgements}
We would like to thank Johan Alwall, Pierre Artoisenet, Pushpa Bhat,
Qinghong Cao, Johannes Heinonen, Wai-Yee Keung, Jen Kile, Andrew Kobach, Andrew
Kubik, Tom LeCompte, Joe Lykken, Olivier Mattelaer, Frank Petriello, 
Seth Quackenbush, Heidi Schellman, Michael Schmitt, 
Shashank Shalgar, Gabe Shaughnessy, 
Tim Tait, and Nhan Tran for useful conversations and/or correspondence.
RVM acknowledges the support of a GAANN fellowship.
This work was supported in part by the U.S. Department of Energy under
contract numbers DE-AC02-06CH11357 and DE-FG02-91ER40684.
\end{acknowledgements}

\end{document}